# Experiments demonstrating the creation of elements via pulsed electric field in water and deuterated water: application to the production of fluorine

Gordon Ross,[a] Dominic Anwar,[a] Daniel Sedgwick,[b] Santos Fustero[b]

[a] Team Scotia Ltd, Glenruthven Mill, Abbey Road, Auchterarder, Perth & Kinross, Scotland

[b] Departament de Química Orgánica, Universitat de València, 46100 Burjassot, Valencia, Spain
E-mail: gordie@teamscotia.com
E-mail: santos.fustero@uv.es

## Abstract

A relatively simple experiment, which has been proven to be fully reproducible, shows that in solutions of water and deuterated water, sensible quantities of fluorine are generated when a triggering energy is supplied in form of electric field with suitable characteristics leading to discharges in the material between the electrodes. The novelty of the experiment is the simplicity of the layout, the full reproducibility and the concrete proof of the events based on simple techniques to detect elements that, as it was accurately checked, were not present in the system before energy supply. Modularity is another interesting feature of the solution presented here.
The experimental apparatus used (named STORM REACTOR® and described here for the first time in scientific literature) consists of reaction chamber, control system and instrumentation to detect fluorine. It is illustrated together with the measured results and with discussions on the issue of reproducibility and on possible application areas, including energy production. The case of fluorine production is of potential major importance at industrial level being fluorine widely employed in medical, materials and agrochemical sciences (in particular artificial radioactive fluorine isotopes in medical diagnosis). Moreover, its production with traditional methods can be considered difficult and expensive as compared to generation in a safe and inexpensive way using the device presented here.

## 1. Introduction

A variety of activities have been going on for decades on the subject of inducing nuclear fusion via 'alternative', rather than conventional methods. For example, on the topic of Low Energy Nuclear Reaction (LENR) a large number of papers have been published [1] and are being published in particular in journals such as Journal of Condensed Matter Nuclear Science [2] and a dedicated international Congress is devoted to this theme, every year. A Community funded research Program is going on [3]. As an outcome of these activities, a series of experimental results on transformation of matter, accompanied in some cases by a possible generation of energy, are well documented in scientific literature – a selection thereof can found in [4] but it must be acknowledged that these give rise to undeniable contradiction with established physical theories and have not been interpreted yet through models that have reached a widespread consensus. An attempt to indicate a possible line of interpretation has been recently given in [5] where it is described how a theoretical formalism based on Deformed Minkowski Space Time [6] might be useful to explain experimental outcomes of this kind.

The aim of this paper is limited to reporting the results of experiments conducted both at University of Valencia and in the Labs. of Team Scotia Ltd. . These results lead to the conclusion that, through a process which is yet to be explained by mainstream science, elements in a reactor chamber in the presence of water and pressurized air under proper voltage stimulation, are transformed into atoms of fluorine that were not present before the discharges. The evidences emerging from the experiments indicate that energy is also produced. Nevertheless, the paper focuses on fluorine production and possible uses since fluorine chemistry is gaining more and more interest over a variety of scientific fields such as in pharmaceuticals, agro-chemistry and material science [7]. This is due to the unique properties that fluorine bestows upon molecules that contain it; since it is the most electronegative atom, the C—F bond is highly polarized and has a certain ionic quality [8]. Of particular importance is the use of artificially produced radioactive isotopes $^{17}$F and $^{18}$F for several applications including medical diagnostics.

## 2. Description of the experimental apparatus

The general layout of the equipment which was used and patented with the name STORM REACTOR® [9] , is outlined in Figure 1.

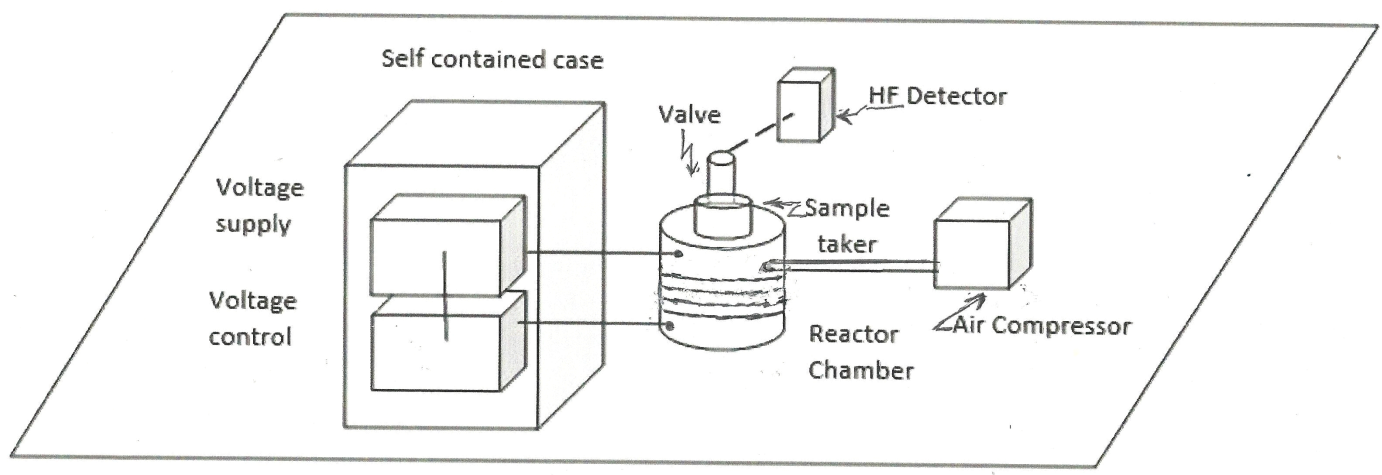

Figure 1. The general layout of the equipment

### 2.1 The reaction chamber

The reaction chamber consists of a container made of stainless steel partially filled with water and pressurized, before providing triggering voltage, in a range from 30 psi to 50 psi through an air compressor). The reaction chamber is surrounded by a copper solenoidal coil which has proven to be useful to trigger the onset of discharges.

In the container, having a volume of approximately 50 cubic centimeters, two electrodes are located; the discharges produced when voltage to the electrode is provided give place to extremely short lived bursts of pressure.

2.2 The voltage supply and control system

The reactor is activated and controlled via an electrical circuit containing a plurality of electrodes, capacitors and inductors, as shown in Figure 2, in a self-contained case. By means of this circuit electric discharges are produced in the reaction chamber at high operating frequency, using two voltage supplies, a DC supply at 2000V and a second AC supply which starts at 40,000V. Electric power is also supplied to the solenoid coil surrounding the reactor chamber which is in series with the electrodes contributing to the total inductance of the circuit.

## 3. Measurements and results

In the present paper, which deals with experiments conducted at University of Valencia and in the Labs. of Team Scotia Ltd., the attention is focused on the experimental evidence that fluorine is produced. The issue of excess heat generation is addressed here only by mentioning clues that this effect appears to be an outcome of the experiments.

3.1 Procedures

$H_2O$ was placed between the electrodes; the electrodes are then attached to the reactor chamber via a screw fitment. The air valve assembly was opened and the chamber compressed to a selected pressure. Once the reaction chamber reached the desired pressure, the air valves were closed and the airline removed. The fluorine meter was connected to the valve assembly whilst the valve remained closed (to expedite the measurement of fluorine, following a reaction). The reading of the fluorine detector showed at this stage no fluorine detection.

The electric circuit was charged through the DC voltage supplier. Then, through the AC voltage supplier, operated either as a manual sequence of single shots or via a signal generator, the discharges were activated. Following the data logging of the reaction, the valve assembly was opened allowing the decompression of the reaction chamber and the fluorine peak value (PPM) measured in the fluorine detector was logged.

3.2 The instrumentation used

To conduct the experiment different combinations of measurement tools were used, besides electrical input power meter, including a Piezoelectric Pressure Transducer, a "Max Pressure" indicator and an oscilloscope measuring voltage and current

To detect the evidences of the consequences of the experiments (besides pressure gauges to measure pressure bursts) a GASTiger2000 HF3 was used to show that fluorine was generated. The HF detector in the Valencia test had a max calibration of 5 ppm and increments of 0.01 ppm. The max calibration was put to 10 ppm in the Scotia tests.

3.3 Results

The activation of the discharge through water and pressurized air, produces well measurable amounts of HF, depending on the type of water used. A total of 40 tests were performed.

The amount of HF produced was measured in air water mixture flowing through the valve to the GASTiger2000 HF detector. Using water samples of different sources (including both water from the tap in different locations in UK and Spain, distilled water, ultrapure water - HPLC grade - and water where NaCl was added) the results obtained in each single activation were that the maximum calibration (5 ppm in one set of tests, 10 ppm in other tests) was exceeded and that the Geiger counter was activated. The circuit could be charged in this configuration with a repetition rate of a few seconds.

When dimethyl sulfoxide DMSO was introduced instead of water the reading on the HF detector was significantly lower (at least a factor 5) supporting the assumption that HF is produced when water is used in the reactor. Furthermore, the reaction took place with a similar efficiency when deuterated water was used, which could be the key to producing $^{18}$F as discussed later.

The intensity of pressure bursts and the quantity of HF produced resulted to be dependent on the characteristics of the AC electric power supplied to the system.

These results lead to the conclusion that, through a fusion process which is not explained yet (the tentative name adopted elsewhere is nuclear metamorphosis) elements in the reactor chamber in the presence of water and pressurized air under proper voltage stimulation, are transformed into elements that were not present before the discharges.

At the stage of experiments reported here it had not been established experimentally yet what isotopes of fluorine are produced. By a simple application of the law about the conservation of the baryonic number it can be inferred that nuclei of hydrogen (in water) and oxygen (in water and in pressurized air), give place to $^{17}$F. Replacing hydrogen with deuterium (when using deuterated water) $^{18}$F should result. Other combinations producing fluorine isotopes can be envisaged considering the involvement of nitrogen (in pressurized air and in particular in air bubble in water) with two nuclei of hydrogen instead of one (analogously with two nuclei of deuterium).

### 3.4 Reproducibility

Lack of reproducibility has been the weak point of most activities reported in literature in the area of LENR. An explanation could be the presence of effects of anisotropy and asymmetry on one side and of a pulse structure versus time on the other side [5]. The experiment described here is not subject to this weakness. The motivation behind could be that an integration over space is ensured by the geometrical layout and a time integration results from measuring the quantity of fluorine accumulated after several bursts.

## 4. Some interesting collateral evidences registered

To overcome any uncertainties it might be useful to mention, as accurately as possible, even minor collateral evidences identified during the tests. These observations could be clues facilitating confirmation (or on the contrary rejection) of interpretation or even explanation of physical phenomena that produce the major effects detected (in this case ascertained production of fluorine and inferred excess heat generation). A list is given in the following.

*Air bubbles:* When an experiment was conducted and the electrodes were removed for observation, the remnant liquid (presumably H2O) was full of gases which were bubbling and escaping as the remaining liquid quickly evaporates.

*Debris:* There were always residuals which build up in powder form on the spark plug. We have not identified these but have hypothesized that they might be the result of nuclear metamorphosis at the transition between the water and the electrode.

*Deformation of electrodes***:** Extensive corrosion of the surface of the electrodes has taken place after the discharges

*Light emissions:* During or following the pressure increase phase, visible light emissions have been witnessed.

*Presence of hysteresis phenomena:* In voltage mapped against the pressure in the measurement chambers huge reductions in pressure were recorded just prior to large voltage and pressure increases. Following the production of HF, there was a long lived change to the resistance of the electrodes when measured with an Ohm meter, which was not present if HF was not produced.

*Pulse nature of the phenomenon*: The heat generation (as signaled by pressure change measurements) seems to occur in a few waves of production before settling.

*Effects of geometry of the layout:* The shape of the reactor chamber as well as the shape and dimensions of the electrode had an effect on the intensity of the phenomenon.

# 5. Possible application areas

## 5.1 Application to fluorine production

Historically fluorine has been notoriously difficult to work with, given that it is the most reactive of all elements and exists as a highly toxic gas and a costly process is still employed for its production. Hydrogen fluoride, commercialized in a variety of forms, is among the most notable and versatile electrophilic and nucleophilic reagents. The only natural isotope of fluorine is $^{19}F$, which is incidentally one of the best tool for use in NMR spectroscopy.

A major economic interest is in the production of unstable artificial isotopes of fluorine. Specifically, $^{18}F$ is widely used as a radiotracer in the form of $Na^{18}F$ or fluorodeoxyglucose. For approximately 80% of the world's nuclear medicine diagnostic imagine procedures, $^{99m}Tc$ (a metastable isotope of Technetium) was originally used. But, since its worldwide shortage of in 2009 [10], radioactive sodium fluoride has emerged as a powerful alternative for bone imaging [11], [12], [13]. Fluorodeoxyglucose, on the other hand, is used to track glucose metabolism in patients; tissues with high metabolic activity, such as malignant tumors, retain fluorodeoxyglucose and their location can therefore be seen through PET imaging [14]. The production of $^{18}F$ (half-life equal to 109.8 minutes) is at present time much more difficult than the production of $^{99}Tc$ m; it can only be done in cyclotrons, vastly limiting the availability of so-called "hot" reagents. The ability to produce $^{18}F$ directly in the medical centre or hospital for immediate use would be of great interest to the scientific and medical community. The use of deuterated water in the reactor chamber could lead to the production of $^{18}F$.

Also $^{17}F$ (which decays by positron emission mode with 96% probability and by electron capture mode with 4% probability) may be a useful compound for medical imaging. However, limitations arising from its production and very short half-life of just 64 seconds have restricted its use. Fluorine-17 may decay giving place to the molecule $H_2[^{17}O]$ which could also be useful, given that it is stable and NMR-active.

The above compounds are to be considered at present as difficult and expensive to produce. Therefore, there is a need for systems and methods that allow for their easier and less expensive production. In this sense, the STORM REACTORs® when developed into an industrial production tool, could represent a novel and sustainable way of accessing fluorinated compounds by producing reactive fluorine species from just water, although the reactivity of the species in the mixture produced is still to be determined. Further tests are currently being carried out to confirm this and to improve the efficiency of HF generation.

## 5.2 Different potential application areas

### 5.2.1 Matter transformation

The presence, when ascertained experimentally, of other nuclei in the reactor chamber as well as of nuclear particles might shed light on the mechanisms. A more complete set of experiments with the involvement of instruments including mass spectrometers gamma ray spectrometry, calorimetric techniques is under way. As a following step, experiments will be conducted adding other nuclides in the reactor chamber in order to investigate the viability of producing isotopes of interest including high value ones - such as rare earths **[15]** - and of reducing the radioactivity of potentially dangerous radioactive materials **[16]**, **[17]**, **[18]** such as nuclear waste.

*5.2.2 Energy production*

A large number of experiments dealing with LENR reported in literature mention indications of energy production in terms of heat produced in excess of the quantity expected as a consequence of electric energy supplied by the electric grid.

This is a very controversial issue; a well-known example is the case of the E-Cat device **[19]**. The authors underline that the use of pulsed voltage stimulation characterizing the experiment reported here is different, as compared with the circulation of current in a water electrolytic solution, and that, having taken into consideration the criticisms formulated on the layout and measurement techniques relative to that experiment, they have deployed countermeasures that can be considered adequate to overcome such criticisms. In particular, it is recalled that at the present stage the authors of the present work consider as fully proven only the production of new fluorine nuclides, which is independent form the critical remarks directed to the E-cat experiment, dealing with the claimed performance of generating excess heat for practical applications.

# 6. Conclusive remarks

6.1 Comparison with other experiments and possible industrial applications

As compared to the large number of experimental configurations set up to investigate LENR, the novelty of the experiment is both in the simplicity of the layout - with full reproducibility - and in the concrete proof of the events which is based on simple techniques to detect elements that, as it was extensively checked, were not present in the system before energy supply. Modularity and compactness are other interesting features of the solution presented here.

The arrangement utilized by Urutskoev et alii **[18]**, **[20]** can be considered as a similar one, since in both cases high voltage discharges are produced in water using metallic electrodes. The major difference is that in the Urutskoev's experiments one of the electrodes is in form of a thin foil and it is expected to be destroyed during the test. In both cases new chemical elements are detected and "strange" radiation associated with the transformation of elements was registered; in the Russian experiments use was made of spectroscopic measurements during the electric discharge and of a mass-spectrometer analysis of sediments after the discharge. The advantage of the configuration adopted in the present paper is the possibility of accumulating the produced isotopes by repeating the discharges (multi shot) instead of being forced to work in a succession of one shot operations with the necessity of replacing the foil each time. Another advantage is the possibility of optimizing the resulting yield by proper modulation of the shape and frequency of the triggering pulses of voltage which is not easy to do with the rigidity of simply switching on-off a battery.

It is not to be underestimated the practical circumstance that there is concrete demand in the market for the fluorine produced by using the STORM REACTOR®.

6.2 Contributions to the development of research on the alternative fusion theme.

The experiments reported here lead to the well proven conclusion that in the reactor chamber in the presence of water and pressurized air, under proper stimulation through electric discharge stimulation, elements that were not present before the discharges are produced systematically in a reproducible manner.

The process underneath these results is not unanimously established yet (one example of a possible theoretical explanation is given in Ref. [5] ). Even if suggesting a comprehensive explanation is not among the goals of this paper, two contributions are given to this purpose:

- a simple device is proposed which is suitable for further investigations and provides fully reproducible results
- collateral evidences identified during the tests are reported that, even if deemed of minor importance, could, on the contrary, be clues facilitating confirmation (or on the contrary rejection) of interpretation or even explanation of physical phenomena.

6.3 Future experimental activities planned

Further experimental activities are under way both at University of Valencia and in the Labs. of Team Scotia Ltd., to investigate systematically the aspect of heat excess generation, in coherence with the goal of the program underway under the auspices of the European Commission and the connected destination of financial resources [3].

## 8. Acknowledgments

The authors thank Stefano Bellucci and Fabio Pistella - two of the authors of the paper mentioned as Ref. [5] - for useful discussions. Daniel Sedgwick, and Santos Fustero are also grateful to the Team Scotia Ltd. and the University of Valencia (UV) for financial support.